\documentclass[prl,aps,twocolumn,epsfig]{revtex4}
\usepackage{epsfig}
\usepackage{amsmath}
\usepackage{amssymb}
\usepackage{bm}
\usepackage{dcolumn}
\usepackage{ulem}

\newcommand{\beq}{\begin{equation}}
\newcommand{\eeq}{\end{equation}}
\newcommand{\bea}{\begin{eqnarray}}
\newcommand{\eea}{\end{eqnarray}}
\newcommand{\bei}{\begin{itemize}}
\newcommand{\eei}{\end{itemize}}

\def\bigout#1{#1$\!\!\!\!\! \backslash$}
\def\smallout#1{#1$\!\!\! \backslash$}

\begin{document}

\title{Breaking of Particle-Hole Symmetry by Landau Level Mixing in the $\nu=\frac{5}{2}$ Quantized Hall State}

\author{Edward H. Rezayi${}^{1}$ and Steven H. Simon${}^{2}$}
\affiliation{${}^1$Department of Physics, California State University, Los Angeles, California 90032\\
${}^2$ Rudolf Peierls Centre for Theoretical Physics, 1 Keble Road, Oxford University, OX1 3NP, UK}

\begin{abstract}
We perform numerical studies to determine if the fractional quantum Hall state observed at filling $\nu=5/2$
is the Moore-Read wavefunction or its particle hole conjugate, the so-called AntiPfaffian.
Using a truncated Hilbert space approach we find that for realistic interactions, including Landau-level
mixing, the ground state remains fully polarized and the the AntiPfaffian is strongly favored.
\end{abstract}
\maketitle

Over the last twenty years one of the most intriguing puzzles in
condensed matter physics has been the nature of an ``even denominator''
quantum Hall state observed\cite{Willett} at filling fraction $\nu=5/2$.  
Shortly after its discovery it was shown\cite{Eisenstein} that the $5/2$ plateau disappears when a
sufficiently strong in-plane magnetic field is applied to the 2-D electron layer. This observation was
widely assumed to indicate a spin-singlet (or partially polarized) ground state, since increasing
the in-plane field would force the spins to align and destroy such a ground state.   However,
numerical results by Morf\cite{Morf} showed that in a relatively large finite size system, the polarized
state is preferred over spin-singlets even in the absence of the Zeeman gap.  There is now agreement among
numerical calculations\cite{Morf,Rezayi00, FurtherNumerical}, using a variety of techniques and in
different
geometries, that the polarized ground state is adiabatically connected to a gapped phase, generally believed to be the Moore-Read\cite{MooreRead}
Pfaffian (Pf) state, but lies very close to a quantum phase transition to a compressible phase so that
small changes in details, such as tilting the magnetic field, can push the system across the phase
boundary as had been observed experimentally.
Despite this convergence of views there remains a serious complication in comparing the Pfaffian state
to the ground states found in numerical studies:  It was recently realized\cite{AntiPfaffian} that
the particle-hole conjugate of the Moore-Read Pfaffian
state, called the AntiPfaffian (APf), is an equally valid candidate for systems which obey particle-hole
symmetry; and no prior numerical study would have had the ability to distinguish between these two
possibilities\cite{Radu}.  On the other hand, the actual experiments do not obey particle-hole symmetry and so
the two candidates are inequivalent in experiments and, most likely, only one is realized.  While
the Pfaffian and the AntiPfaffian share many important properties, including non-Abelian statistics and
their usefulness for quantum information processing\cite{NayakRMP}, they represent different
topological phases of matter. For example, they have different edge physics\cite{AntiPfaffian}.  Determining
which of these two states is actually realized in the physical system is now a rather crucial objective.
This is the aim of the current paper.

In all previous numerical studies, calculations were performed in a Hilbert
space restricted to a single Landau level (LL).  Indeed, it is precisely this
restriction that allows numerical calculations in systems of
relatively large sizes (up to 22  electrons in some cases\cite{FurtherNumerical}).
However, if the inter-particle interaction is two-body, such a  restriction enforces a precise symmetry
between particles and holes  that is
absent in the actual experiments, and this symmetry is broken once LL
mixing is restored.   While for most studies  mixing makes only quantitative (albeit important)
corrections, in the case of 5/2 it is crucial since it is the only factor that selects between
Pfaffian and AntiPfaffian.
Without consideration of LL mixing, the system
is fine-tuned to a critical point between these phases\cite{Sheng}.
Furthermore, single LL projection can only ever be quantitatively accurate to the extent that the
Coulomb energy $E_c = e^2/\epsilon \ell_0$  is much smaller than the cyclotron
energy $\hbar \omega_c$. Unfortunately, in actual quantum Hall experiments
this is never so. In fact $E_c / \hbar \omega_c \gtrsim 1$ in all published
experiments on 5/2.  In the current paper we will set $E_c / \hbar \omega_c =1.38$  as in
Ref.~\onlinecite{Pan}.

The challenge of attacking this problem numerically is that without the
projection to a single LL, the Hilbert space, even for a small system,
is infinite.  One approach to addressing this problem is to integrate out
LL mixing terms perturbatively in $E_c/\hbar \omega_c$, leaving an effective
single LL theory.  This approach has recently been implemented to lowest order
in Ref.~\onlinecite{Chetan}.  The single LL theory has been obtained, and it
was tentatively concluded that at lowest order the LL mixing terms most likely
favor AntiPfaffian over Pfaffian.  While such an approach is well controlled,
it also has obvious severe limitations: it is a perturbative expansion in a
parameter which in the experiment is of order one.  While in principle one
could attempt to continue the expansion to higher orders hoping that the series
converges quickly, even if one could perform the more complicated algebra, at
higher order one generates retarded interaction terms which then makes the
resulting single LL analysis very difficult.

An alternate, seminumerical, organization of a perturbative expansion in LL
mixing terms was developed in Ref.~\onlinecite{HaldaneRezayiLLMixing}.
We attempted a similar approach, hoping that terms of successively higher
orders in $E_c/\hbar \omega_c$ would become rapidly less important.  However,
for $E_c/\hbar \omega_c \gtrsim 1$ corresponding to the experiments,  we found
that one would have to carry out this expansion to a higher order than
feasible in order to obtain convergence.
We have, therefore, resorted to a different method of analysis.

The approach we take is non-perturbative and is similar in spirit to that of Ref.~\onlinecite{Wojs}.
We use a truncated Hilbert space method by keeping a limited number of LL's  and allowing only a certain
number of particle or hole excitations out of the valence LL, and performing exact diagonalizations on
this restricted Hilbert space.   One may view such an approach
as variational in character, which can successively be improved by further expanding the Hilbert space.
We note that matrix elements connecting the valence LL to increasingly high
LL's drop very rapidly, so excluding high LLs is not expected to create
substantial errors (see for example \cite{HaldaneRezayiLLMixing}).

Even given the Hilbert space truncation, it is still challenging to
establish meaningful results from the limited size system.  As such, our
argument will proceed in three steps.  (1) We show that the
valence LL is polarized.   This result was first established without LL
mixing in Ref.~\cite{Morf}, and is re-examined here for completeness as well
as for ascertaining that LL mixing does not change this conclusion.
This then allows us to concentrate on systems with a polarized valence LL.  (2) We establish
that excitations of electrons with opposite spin from that of the valence LL do
not substantially effect the crucial physics.   This allows us to further
reduce the Hilbert space dimension considerably. (3) Finally, using our truncation method
we can meaningfully address adequate system sizes
and accurately determine the nature of the ground state.

We perform our calculations on the torus geometry at $\nu=5/2$ where
the Pfaffian and AntiPfaffian compete with each other directly (as compared to
the sphere where, for a particular number of particles, the Pfaffian and
AntiPfaffian do not occur at the same flux). Unless otherwise stated we
will use a hexagonal unit cell.   Following Haldane\cite{HaldanePBC}, we use two-dimensional conserved crystal
momentum to classify the states.
We consider two different classes of experimental samples where
$\nu=5/2$ has been observed.   The first class, typical of earlier experiments\cite{Willett,Eisenstein},
include single heterointerfaces.  In the current work we will focus on samples of this type using a Fang-Howard (FH) layer profile\cite{Fang}, with a layer width of $w=0.65$ magnetic lengths\cite{Pan}.
The second class of samples is the
somewhat wider (roughly 30 nm) symmetric quantum well (QW) typical of modern high-density ultra-high mobility experiments\cite{Xia}. 
In this case, the LLL of the first excited subband may lie below the
$2\downarrow$ LL of the lowest subband and hence should not be ignored\cite{Eisenstein_PC}.
We will return to the QW case before concluding.

\begin{table}
\begin{tabular}{c||c|c|c|c|c|c|c|}
{\bf (a)} & $N_{\phi}$ & $N$ & 0$\downarrow$+0$\uparrow$ & 1$\downarrow$+1$\uparrow$ & 2$\downarrow$+2$\uparrow$ & 3$\downarrow$+3$\uparrow$ & dim \\
\hline \rule{0pt}{9pt}
${\cal H}_{v,6}$  & 12  &  30  & \bigout{24}  & 6 & \smallout{0} &  \smallout{0} &   6.7$\times10^2$\\
${\cal H}_{v,8}$  & 16  &  40  & \bigout{32}  & 8 & \smallout{0} &  \smallout{0} &    2.6$\times10^4$\\
${\cal H}_{v,10}$  & 20  &  50  & \bigout{40}  & 10 & \smallout{0} &  \smallout{0} &    1.2$\times10^6$ \\
${\cal H}_{s,1}$  & 12  &  30  & 22-24  & 6-8 & \smallout{0} &  \smallout{0} &   7.6$\times10^4$\\
${\cal H}_{s,2}$  & 12  &  30  & 23-24  & 5-7 & 0-1 &  \smallout{0} &   2.5$\times10^4$\\
${\cal H}_{s,3}$  & 12  &  30  & 22-24  & 5-8 & 0-1 &  \smallout{0} &   1.1$\times10^6$\\
${\cal H}_{s,4}$  & 12  &  30  & 22-24  & 4-8 & 0-2 &  \smallout{0} &   6.0$\times10^6$\\
${\cal H}_{s,5}$  & 16  &  40  & 31-32  & 7-9 & 0-1 &  \smallout{0} &   4.5$\times10^5$\\
\end{tabular}

\vspace*{10pt}

\begin{tabular}{c||c|c|c|c|c|c|c|c|c|c|}
{\bf (b)} & $N_{\phi}$ & $N$ & 0$\downarrow$ &  1$\downarrow$ & 2$\downarrow$ & dim & Pf & APf & $\, \langle {\rm P} | {\rm A} \rangle^2$\\
\hline \rule{0pt}{9pt}
${\cal H}_{p,6}$  & 12  &  30  & \bigout{12} &   6 & \smallout{0}   & 14 & .90 & .90 & .69 \\
${\cal H}_{p,8}$  & 16  &  40  & \bigout{12} &  8 & \smallout{0}  & 1.0$\times10^2$ & .53 & .53 & .016 \\
${\cal H}_{p,10}$  & 20  &  50  & \bigout{12} &  10 & \smallout{0}  & 9.2$\times10^2$ & .71 & .71  & .29 \\
${\cal H}_{p,12}$  & 24  &  60  &  \bigout{12} & 12 &  \smallout{0}  & 9.4$\times10^3$ & .56 & .56 & .059 \\
${\cal H}_{r,1}$  & 12  &  30  & 10-12 &   6-8 &  \smallout{0}   & 6.0$\times10^2$ & .94 & .83 & .69\\
${\cal H}_{r,2}$  & 12  &  30  & 10-12 & 5-8 &  0-1  &   1.1$\times10^4$ & .80 & .89 & .69 \\
${\cal H}_{r,3}$  & 12  &  30  & 10-12 &  4-8 &  0-2  &  7.6$\times10^4$ & .83 & .89 & .69 \\
${\cal H}_{r,4}$  & 12  &  30  & 9-12 &  3-9 &  0-3  &  1.1$\times10^6$ & .82 & .89 & .69  \\
${\cal H}_{r,5}$  & 16  &  40  & 14-16 &  8-10 &   \smallout{0} & 9.1$\times10^3$ &.63  & .33 & .016\\
${\cal H}_{r,6}$  & 16  &  40  & 14-16 &  7-10 &  0-1 &  2.1$\times10^5$ & .34  & .51 & .016 \\
${\cal H}_{r,7}$  & 16  &  40  & 14-16 &  6-10 &  0-2 &  1.8$\times10^6$ & .37 & .56  & .016\\
${\cal H}_{r,8}$  & 20  &  50  & 18-20 &  10-12  & \smallout{0}   & 1.4$\times10^5$& .40  & .00 & .29 \\
${\cal H}_{r,9}$  & 20  &  50  & 18-20 &  9-12  & 0-1   & 3.8$\times10^6$ & .01  & .24  & .29 \\
\end{tabular}

\caption{Hilbert Spaces.   The first column is the
label of Hilbert space used in the text.  The 2nd column is total flux $N_\phi$,
the 3rd column $N=(5/2) N_{\phi}$
is the total number of electrons in the system.  The next columns are the
number of electrons allowed  in each LL respectively (with $\uparrow$ and
$\downarrow$ representing spin up and down LLs).   Entries with a slash
through them indicate that there is no freedom in that LL (i.e., it is
frozen as entirely filled or empty).    In  table (a) only the total
number of spin up plus spin down electrons in a LL is fixed.
In table (b)
the spin up particles are always frozen (0$\uparrow$ is completely filled
with $N_\phi$ electrons and all other $\uparrow$ LL's are empty) and the
number of electrons in spin down LL's are specified.  Higher LL's are
assumed empty.  The column labeled ``dim'' shows the crystal momentum
reduced dimension of the Hilbert spaces.  For (b), the
projected square overlap of the exact ground state
with the Pfaffian and AntiPfaffian trial states are shown
in the  columns labeled Pf and APf respectively.  With sufficient LL mixing
in almost every case, particularly for the
larger systems,
the AntiPfaffian is favored.    Furthermore, all the APf overlaps increase as $V_1$ is increased slightly
(see Figs \ref{fig:overH}, \ref{fig:overS}).  The final column is the square overlap of Pfaffian with
AntiPfaffian for the given system size.}
\label{tab:hilbert}
\end{table}

Some of the Hilbert spaces which we examine are listed in Table
\ref{tab:hilbert}. The simplest of the Hilbert spaces are
${\cal H}_{p,{\cal N}}$ which describe ${\cal N}$ electrons filling half
of the 1st spin down LL (where all other LL's are either completely
filled or completely empty).   This type of space, where all of the
degrees of freedom are within one spin polarized LL, is typically the space
used for study of the quantum Hall effect. Indeed, the trial wavefunctions we
are interested in comparing to, the Pfaffian and the AntiPfaffian, are
completely contained within this space.  It is useful to define a normalized
projection operator \begin{equation}\hat P_{p,\cal N} | \psi \rangle
= P_{p,\cal N}| \psi \rangle \, / \, |\langle \psi| P_{p,\cal N} | \psi
\rangle|^{1/2}\end{equation}
where $P_{p,\cal N}$ is the usual projection to the Hilbert space
${\cal H}_{p,{\cal N}}$.  Thus, $\hat P_{p,\cal N} | \psi \rangle$ is
always a normalized wavefunction within the space ${\cal H}_{p,{\cal N}}$.
We can further define the projected square overlap  between two states
$\psi_1$ and $\psi_2$ with the same total number of electrons $N$
as $|\langle \psi_1 |\hat P_{p,\cal N} \hat P_{p,\cal N} | \psi_2
\rangle|^2$ where ${\cal N} = N \bmod {N_\phi} =N/5$.

We start by examining spin polarization of the valence LL.
First we  restrict the Hilbert space to a single LL (${\cal H}_{v,i}$ for $i=6-10$) and, before doing exact diagonalization,
we integrate out inter-LL transitions approximately
at the RPA level\cite{AleinerGlazman}, which modifies the inter-electron interactions.
We find, in agreement with Ref.~\onlinecite{Morf}, that even in the absence of Zeeman energy,
the ground state of
the valence LL is fully polarized and gapped for ${\cal N}=6-10$ electrons. Furthermore the signature
crystal momentum of the Pfaffian state (and the AntiPfaffian) for both even and odd electrons is matched by the
exact ground state.
We find these results to be true for both FH and QW layer profiles\cite{unpublished}.

\begin{figure}
\vspace*{-30pt}
 \begin{center}
 \includegraphics[width=1.05\columnwidth]{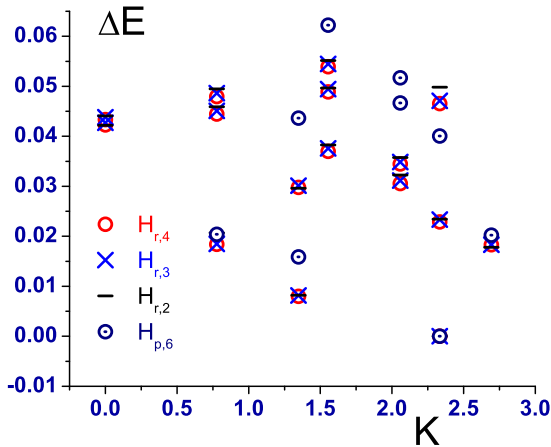}
 \end{center}
 \vspace*{-30pt}
 \caption{(color online) Spectrum of low energy spin-polarized excitations for
   $N=30$ and $N_{\phi}=12$ with Fang-Howard type electron interaction (see text).
   As the Hilbert space is expanded from ${\cal H}_{r,2}$ to ${\cal H}_{r,3}$ to ${\cal H}_{r,4}$, the spectrum converges very quickly.  Also shown is ${\cal H}_{p,6}$ where inter-LL transitions are forbidden, which has a somewhat different excitation spectrum.
 }
 \label{fig:spec2}
 \end{figure}

We reconsider the same problem including LL mixing with the truncated Hilbert space technique. Performing
diagonalizations in Hilbert spaces
${\cal H}_{s,i}$ we now allow complete freedom within the valence LL, and we
also allow a few holes in the 0th LL and a few electrons in the 2nd LL.
We again find that the ground state of the valence LL is
always fully polarized even in the absence of Zeeman splitting and we again find the ground state momentum to match that of the Pfaffian and AntiPfaffian.

Concluding that the valence LL is polarized,
we now turn to study the effect of the excitation of the spin up (the minority spin) electrons.  Let us consider the ground
states ${\cal H}_{r,i}$, which are exactly analogous to ${\cal H}_{s,i}$
except that the minority spin species have now been frozen (we do not allow
any excitations of this species) although LL transitions for the majority
spin are still allowed. We find, rather remarkably, that the projected
squared overlap between the ground state in ${\cal H}_{r,i}$ and the
ground state in ${\cal H}_{s,i}$ is  .9893, .9984, .9986, and .9864 for $i=$2,3,4  and 5
respectively.   Considering the moderate size of the Hilbert space
even after projection to ${\cal H}_{p,8}$ (dim=34 after all symmetries are removed) this result is
significant.   What this means is that, although the ground state wavefunction
is dressed with virtual excitations to the other LLs,
when it is projected back into a single LL, {\it the
wavefunction is nearly independent of whether transitions of the
minority spin species are allowed}.  We note that this insensitivity to
the presence of the minority spins seems to hold independent of our truncation
scheme and details of the Hamiltonian, presumably,
so long as we have a polarized and gapped ground state.
The fraction of the wavefunction that survives
projection is clearly reduced when the Hilbert space is expanded,
but this is unimportant in determining the phase of matter represented by the wavefunction.

The above result now allows us to completely freeze the minority spin species and study larger systems. We now examine a number of different truncation combinations, some of which are shown in Table
\ref{tab:hilbert}.b.   Fig.~\ref{fig:spec2} examines how the calculated (spin-polarized) excitation
spectrum changes as a function of the Hilbert space truncation.  As the Hilbert space is expanded,
the spectrum rapidly converges.   This further indicates that only a few excitations out of the valence
LL need be considered in order to capture the essential physics.

\begin{figure}
 \vspace*{-30pt}
 \begin{center}
 \includegraphics[width=1.05\columnwidth]{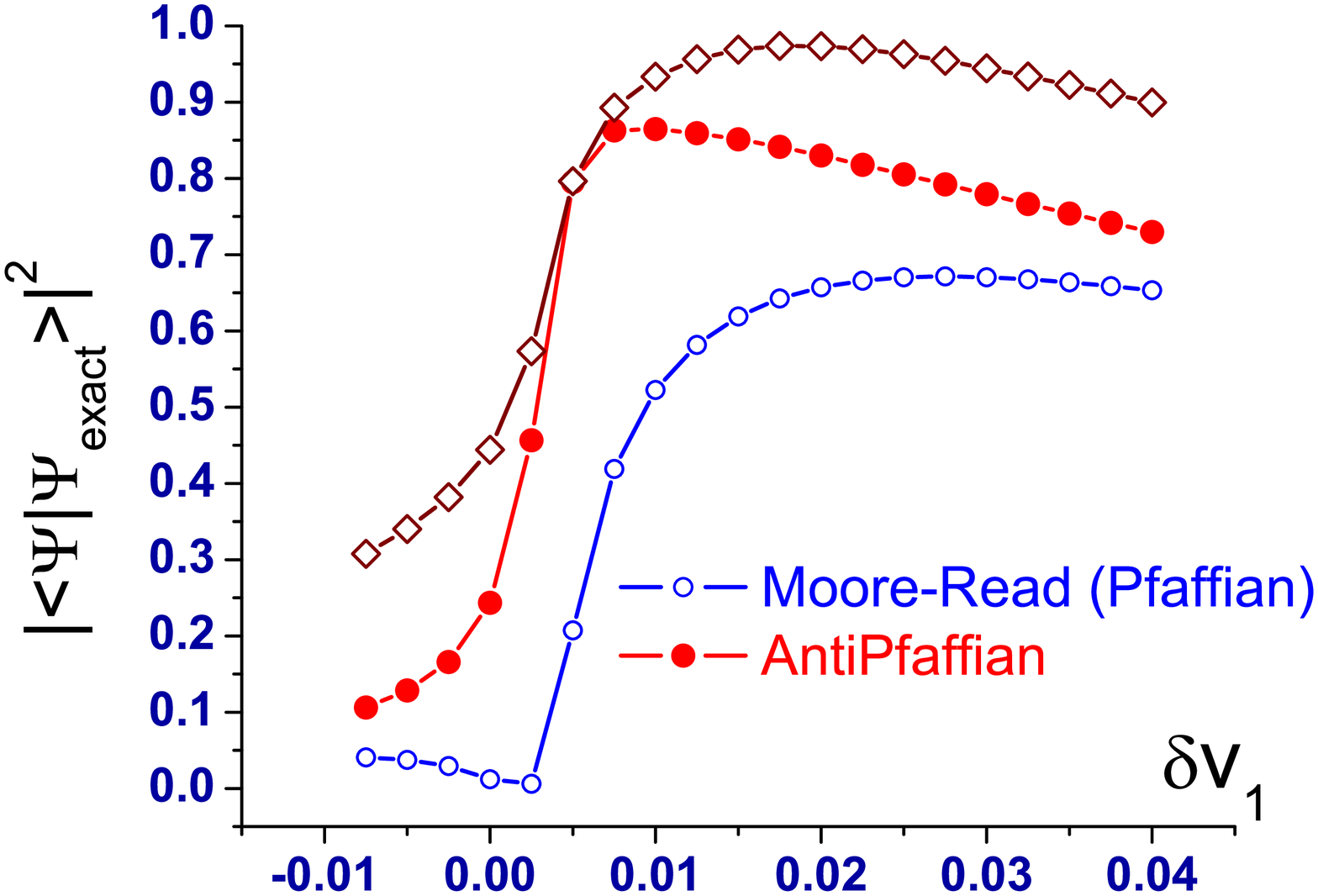}
 \end{center}
 \vspace*{-25pt}
 \caption{(color online) The overlaps of the projected ground state with the MR
Pfaffian and the AntiPfaffian. The system size is
$N=50$ electrons in $N_{\phi}=20$ flux quanta.   Here the Hilbert space used is
${\cal H}_{r,9}$ which allows only spin-polarized LL transitions.   We see that by varying
$V_1$ slightly we find that the ground state has a large overlap with
the  AntiPfaffian, while the overlap with the Pfaffian remains relatively small.
As $V_1$ is increased the two overlaps become comparable.  A large part of this increase
appears to result from a sizable overlap of the Pfaffian and the AntiPfaffian. 
 Further increasing $V_1$, the system  will cross over to the Composite Fermion (CF) Fermi-liquid-like
state\cite{Rezayi00},
which in this case has the same crystal momentum. The top curve (also in Fig.~\ref{fig:overS} for the
ZB case) is the projection of the ground state
into the two-dimensional subspace spanned by the Pfaffian and Anti-Pfaffian.}
 \label{fig:overH}
 \end{figure}

We now turn to the main results of our work.   For increasing system sizes, we consider the projected
overlap of our exact diagonalizations with both the Pfaffian and the AntiPfaffian
(see Table~\ref{tab:hilbert}.b).   Note that for certain finite sized systems, there can be a substantial
overlap between the Pfaffian and AntiPfaffian (see Table~\ref{tab:hilbert}.b), so that if the overlap of
the ground state with one is large, the overlap with the other cannot be too small.
Examining Table~\ref{tab:hilbert}.b, it is clear that the AntiPfaffian is favored over the Pfaffian,
particularly for the larger Hilbert spaces.   Examining the data more carefully makes the case even more
compelling.  For finite sized
systems, the realistic Hamiltonian is clearly on the edge of a phase transition, in agreement
with experiment\cite{Eisenstein} and prior numerical studies\cite{Morf,Rezayi00,FurtherNumerical}.  However, barring a level crossing transition, which does not occur here, phase boundaries are broadened in
finite sized systems so in order to focus on the gapped phase, we add a small $\delta V_1$
(Haldane pseudopotential) interaction to the Hamiltonian.  We define $V_m$ to be the the energy
of a pair of particles in a state of ``relative angular momentum" $m$ irrespective of what LL
they occupy\cite{HaldanePBC}. Defined this way, such a term
does not break particle-hole symmetry.  In Fig.~\ref{fig:overH}, in a system
with a very large Hilbert space, we see that adding only a small $\delta V_1$ greatly increases the
projected overlap of the ground state with the AntiPfaffian trial state, but  to a much lesser degree
with the Pfaffian.   Since for this size system, the overlap of the Pfaffian and AntiPfaffian
is about $29\%$, much of the increase for Pfaffian appears to be caused by this effect.

\begin{figure}
 \vspace*{-30pt}
 \begin{center}
 \includegraphics[width=1.05\columnwidth]{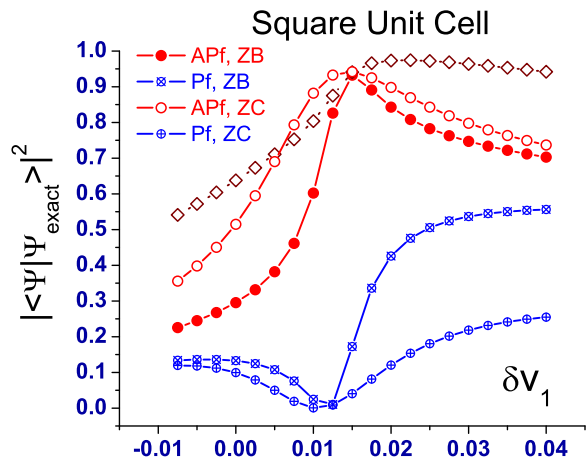}
 \end{center}
 \vspace*{-25pt}
 \caption{(color online) Same as Fig.~\ref{fig:overH} but for a square periodic unit cell. ZC and ZB refer
to the zone
corner and boundary of the Brillouin zone.  Note that for this geometry the overlap of the Pfaffian
with the AntiPfaffian at the ZC is extremely small. A first order transition to the CF state occurs for
$\delta V_1 > 0.04$.  For negative $\delta V_1$ we expect a stripe-type order\cite{Rezayi00}.  Here,
as well as in Fig.~\ref{fig:overH}, the completely symmetry reduced Hilbert space dimension within ${\cal H}_{p,10}$ is 263.}

 \label{fig:overS}
 \end{figure}

It is useful to examine also the torus with square unit cell geometry where additional information may be
extracted  (See Fig.~\ref{fig:overS}).  In contrast to the hexagonal unit cell, where there is a single
three-fold degenerate ground state, for the square unit cell we find two low energy ``ground'' states
at two different points in the Brillouin zone: one at the zone corner (ZC) and a doubly degenerate 
ground state at the zone boundary (ZB),
as expected for either the Pfaffian or AntiPfaffian.   In this case, however, the overlap between
the Pfaffian and AntiPfaffian trial wavefunctions are 0.8\% for ZC and 12\% for ZB.   Here, the
contrast between Pfaffian and AntiPfaffian overlaps is even more apparent: examining the ZC we find a
region of $\delta V_1$ where the overlap with the AntiPfaffian is very high, but the overlap with
the Pfaffian is near zero.

At the peak in both figures, the
AntiPfaffian completely dominates the ground state while past the peak both the Pfaffian and
the AntiPfaffian are present and occupy up to 97.5\% of the ground state with the AntiPfaffian the
dominant component (see the top curves).  
Strictly speaking such an admixture of Pf and APf will not occur in a thermodynamic
system since they represent distinct phases of matter.

We return now to the QW type samples.  In the high density (very high mobility) cases,
the LLL of the first excited subband state is about
30\% of $\hbar\omega_c$ above the $n=1$ Landau level. We find that
in a 3-LL model the mixing is somewhat suppressed and the Pfaffian is preffered.
However, adding the fourth ($n=2$ lowest subband)  LL changes it to the AntiPfaffian.  We have tested
convergence when more LLs are added. In particular, adding the fifth ($n=1$, first excited subband)
LL in  a small, $N=30$ and  $N_\phi=12$,
system we have found that overlap changes are less than one quarter of a percent.  
While the sizes we can access in this case are more limited, the case for the AntiPfaffian remains 
relatively strong.  More details will be given elsewhere\cite{unpublished}.

To conclude, we find the AntiPfaffian is strongly preferred.  The large overlaps with the Pfaffian appear to be a finite size effect at least partially due to the
relatively large overlap of the Pfaffian with the AntiPfaffian.

\acknowledgments
The authors acknowledge helpful discussions with J.~Eisenstein, C.~Nayak, W.~Bishara, N.~Read,
F.~D.~M.~Haldane, and W. Pan.   E.~R. is supported by DOE grant DE-SC0002140.


\begin{thebibliography}{50}


\bibitem{Willett} R. L. Willett, et al, Phys. Rev. Lett., {\bf 59} 1776,     (1987).

\bibitem{Eisenstein}
J. P. Eisenstein,  et al, Phys. Rev. Lett. {\bf 61}, 997 (1988)
\bibitem{Morf} R. H. Morf, Phys. Rev. Lett. {\bf 80},1505 (1998).

\bibitem{Rezayi00}
E.~H.~Rezayi and F.~D.~M.~Haldane, Phys.~Rev.~Lett. {\bf 84}, 4685 (2000).

\bibitem{FurtherNumerical} A. E. Feiguin, et al. Phys. Rev. Lett. {\bf 100}, 166803 (2008);
Phys. Rev. B 79, 115322 (2009); G. M\"oller and S. H. Simon, Phys. Rev. {\bf B77}, 075319. (2008);
O. S. Zozulya et al, Phys.  Rev. {\bf B79}, 045409 (2009).

\bibitem{MooreRead}  G. Moore and N. Read, Nucl. Phys. {\bf B360},362 (1991).

\bibitem{AntiPfaffian} M.~Levin, B.~I.~Halperin and B.~Rosenow, Phys.~Rev.~Lett.~{\bf 99}, 236806 (2007); S.-S.~Lee, S.~Ryu, C.~Nayak and M.~P.~A.~Fisher, Phys.~Rev.~Lett. {\bf 99}, 236807 (2007).

\bibitem{Radu}   Of the experiments performed on the 5/2 state, only the work of
I. P. Radu, et al Science {\bf 320}, 899 (2008) has the ability to distinguish between the Pfaffian
and the AntiPfaffian.  These experiments suggest the presence of the AntiPfaffian, consistent with our calculations.


\bibitem{NayakRMP} A. Yu. Kitaev,
Ann. Phys. 303 2 (2003);  C. Nayak et al, Rev. Mod. Phys. {\bf 80} 1083 (2008).


\bibitem{Sheng} H. Wang, et al arXiv:0905.3589; M. R. Peterson et al, Phys. Rev. Lett. {\bf 101},
156803 (2008).

\bibitem{Pan}  W. Pan et al, Phys. Rev. Lett. {\bf 83}, 3530 (1999).



\bibitem{HaldaneRezayiLLMixing} E. H. Rezayi and F. D. M. Haldane, Phys. Rev. {\bf B42} 4532 (1990).

\bibitem{Chetan}  W.~Bishara and C.~Nayak, Phys. Rev. {\bf B80}, 121302(R) 
(2009).


\bibitem{Wojs} A. Wojs, and J. Quinn, Phys. Rev. {\bf B74}, 235319 (2006).


\bibitem{HaldanePBC} F.~D.~M. Haldane, Phys. Rev. Lett. 55, 2095 (1985); F.~D.~M. Haldane, in R. Prange and S. M. Girvin eds, {\it The Quantum Hall effect},
Springer-Verlag, (New York, 1990)


\bibitem{Fang} F. Fang  and W. Howard  Phys. Rev. Lett. {\bf 16} 797 (1966).

\bibitem{Xia}  J. S. Xia et al, Phys. Rev. Lett. {\bf 93}, 176809 (2004).

\bibitem{Eisenstein_PC}  We are grateful to J.P. Eisenstein for pointng out the importance of
the excited subband  levels to us; see Jing Xia and J. P. Eisenstein to be published.

\bibitem{AleinerGlazman} I. L. Aleiner and L. I. Glazman, Phys. Rev. {\bf B52}, 11296 (1995);
R. Morf and N. d'Ambrumenil Phys. Rev. {\bf B68}, 113309 (2003).

\bibitem{unpublished} E.~H.~Rezayi and S.~H.~Simon, to be published.



\end{thebibliography}
\end{document}